\documentclass[twocolumn]{aastex631}

\usepackage{amsmath}
\usepackage{amssymb}
\usepackage{amsthm}
\usepackage{multirow}
\usepackage{mathrsfs}
\usepackage{hyperref}
\usepackage{nameref}

\begin{document}

\title{
%\sout{Differential Emission Measure Analysis of the X-ray Gas in SN 1987A from 2007 to 2021: The Fading Ring and the Brightening Ejecta}\\
Unusual X-ray Oxygen Line Ratios of SN 1987A Arising From the Absorption of Galactic Hot Interstellar Medium\\
%%%\lsun{Any suggestions for a better title?}
}

	\correspondingauthor{Lei Sun}
	\email{l.sun@nju.edu.cn}
	
	\author[0000-0001-9671-905X]{Lei Sun}
	\affiliation{Department of Astronomy, Nanjing University, Nanjing 210023, People's Republic of China}
	\affiliation{Key Laboratory of Modern Astronomy and Astrophysics, Nanjing University, Ministry of Education, People's Republic of China}
    \affiliation{Anton Pannekoek Institute \& GRAPPA, University of Amsterdam, PO Box 94249, 1090 GE Amsterdam, The Netherlands}

    \author[0000-0003-2836-540X]{Salvatore Orlando}
    \affiliation{INAF-Osservatorio Astronomico di Palermo, Piazza del Parlamento 1, 90134 Palermo, Italy}

    \author[0000-0001-5792-0690]{Emanuele Greco}
    \affiliation{Anton Pannekoek Institute \& GRAPPA, University of Amsterdam, PO Box 94249, 1090 GE Amsterdam, The Netherlands}
    \affiliation{INAF-Osservatorio Astronomico di Palermo, Piazza del Parlamento 1, 90134 Palermo, Italy}
    \affiliation{Universit\`a degli Studi di Palermo, Dipartimento di Fisica e Chimica E. Segr\`e, Piazza del Parlamento 1, 90134 Palermo, Italy}

    \author[0000-0003-0876-8391]{Marco Miceli}
    \affiliation{Universit\`a degli Studi di Palermo, Dipartimento di Fisica e Chimica E. Segr\`e, Piazza del Parlamento 1, 90134 Palermo, Italy}
    \affiliation{INAF-Osservatorio Astronomico di Palermo, Piazza del Parlamento 1, 90134 Palermo, Italy}

    \author[0000-0001-7571-2318]{Yiping Li}
    \affiliation{Department of Astronomy, Nanjing University, Nanjing 210023, People's Republic of China}

    \author[0000-0002-4753-2798]{Yang Chen}
    \affiliation{Department of Astronomy, Nanjing University, Nanjing 210023, People's Republic of China}
    \affiliation{Key Laboratory of Modern Astronomy and Astrophysics, Nanjing University, Ministry of Education, People's Republic of China}

    \author[0000-0002-4708-4219]{Jacco Vink}
    \affiliation{Anton Pannekoek Institute \& GRAPPA, University of Amsterdam, PO Box 94249, 1090 GE Amsterdam, The Netherlands}

    \author[0000-0002-5683-822X]{Ping Zhou}
    \affiliation{Department of Astronomy, Nanjing University, Nanjing 210023, People's Republic of China}
	\affiliation{Key Laboratory of Modern Astronomy and Astrophysics, Nanjing University, Ministry of Education, People's Republic of China}



\begin{abstract}

{Recent high-resolution X-ray spectroscopic studies have revealed unusual oxygen line ratios, such as the high O VII forbidden-to-resonance ratio, in several supernova remnants. While the physical origin {is} still under debate, for most of them, {it has been suggested that this phenomenon arises from either} charge exchange (CX) or resonant scattering (RS). In this work, we report the high O VII G-ratio ($\gtrsim1$) and high O VIII Ly$\beta$/Ly$\alpha$ ratio ($\gtrsim0.2$) found {in multiepoch XMM-Newton RGS observations of SN 1987A}. The line ratios cannot be fully explained by non-equilibrium ionization effects, CX, or RS. We suggest the absorption of foreground hot gas as the most likely origin, which plays the major role in modifying line fluxes and line ratios. Based on this scenario, we introduced two Gaussian absorption components at the O VII resonance line and the O VIII Ly$\alpha$ line and constrained the optical depth of the two lines as $\tau_{\rm OVII}\sim0.6$ and $\tau_{\rm OVIII}\sim0.2$. We estimated the temperature as $kT_{\rm e}\sim{0.15}$\,keV and the oxygen column density as $N_{\rm O}\sim{0.5}\times10^{16}$\,cm$^{-2}$ for the absorbing gas, which is consistent with the hot interstellar medium in the Galactic halo. {Neglecting} this absorption component may lead to an underestimation {of} the O abundance. We revised the O abundance of SN 1987A, which is increased by $\sim20\%$ compared with previous results. The N/O ratio by number of atoms is revised to be $\sim1.2$.}

\end{abstract}



\section{Introduction} \label{sec:intro}

%%%\lsun{to be finished}
{High-resolution X-ray spectroscopy of supernova remnants (SNRs) has provided us with profound insights into the radiation mechanisms and chemical compositions of the shocked ejecta, circumstellar material (CSM), and interstellar medium (ISM), significantly advancing our understanding of shock physics, progenitor type, explosion mechanism, and remnant evolution \citep[see, e.g.,][for a recent review]{2023arXiv230213775K}. Particularly, the {emission} line ratios of He-like and H-like ions, such as the O VII G-ratio and the O VIII Ly$\beta$/Ly$\alpha$ ratio, play an important role in the diagnostics on the temperature and ionization age of hot plasmas \citep[e.g.,][]{2010SSRv..157..103P}. However, due to the complex nature of the SNRs, these line ratios can be affected by different physical processes. 

Recently, unusually high O VII G-ratios (or forbidden-to-resonance line ratios) have been found in several SNRs, such {as the} Cygnus Loop \citep{2011ApJ...730...24K,2015MNRAS.449.1340R,2019ApJ...871..234U}, Puppis A \citep{2012ApJ...756...49K}, N49 \citep{2020ApJ...897...12A}, N132D \citep{2020ApJ...900...39S}, G296.1$-$0.5 \citep{2022ApJ...933..101T}, and J0453.6$-$6829 \citep{2022PASJ...74..757K}. While the physical origin {is} still under debate, two major {alternative} scenarios have been proposed, namely {either} the charge exchange (CX) {or} the resonant scattering (RS). 

In the CX process, an electron is transferred from one atom or ion to another. The {newly captured electron} will cascade from an highly-excited state to the ground level and result in a series of emission lines. CX will boost the forbidden line of the He-like triplet, thus lead to a large G-ratio \citep[e.g.,][]{2012AN....333..301S,2016A&A...588A..52G,2023arXiv230111335G}. On the other hand, RS will change the directions of the resonance photons, which may suppress the observed resonance line flux in some circumstances {depending on geometry} and result in a large G-ratio \citep[e.g.,][]{2024ApJ...967...99L}. 

One of the other possible scenarios is the resonant absorption of the foreground hot gas, which has not yet been {detailedly} investigated in the context of SNRs. A hot ISM component with a temperature as high as $\sim0.15$--$0.22$\,keV has been found in the Galactic halo based on its X-ray emission \citep[e.g.,][]{2013ApJ...773...92H,2022PASJ...74.1396U,2023A&A...674A.195P}. The absorption features left by this hot Galactic ISM, such {as} the O VII and O VIII absorption lines, have also been detected in the X-ray spectra of several Galactic and Magellanic X-ray binaries \citep[e.g.,][]{2005ApJ...635..386W,2008ApJ...672L..21Y,2018ApJS..235...28L}. Therefore, it is very likely that the observed O line fluxes of SNRs, especially the Magellanic SNRs, {are} affected by this absorption component in {the} same way.

The study on O line ratios can help us to further explore the role that CX, RS, and hot gas absorption played in the {X-ray spectral properties} of SNRs. More importantly, if these effects have not been properly considered in the X-ray analysis, the measured metal abundances of the shocked ejecta/CSM/ISM may be distorted, and thus any further discussions based on that may be questioned.
%\lsun{response to EG ``- {\it Correct me if I am wrong, if one tries to fit the global absorption (i.e. the sum of all the contribution to absorption) with only one simple absorption component, one may find a value of nH higher than the true one, and potentially lead to an overestimate on the distance of the source. I guess this effect is small for 87A being in the LMC but may be worth to mention for a generic case}'':
%I believe you are right --- at least from the qualitative point of view. However, I have another concern about this issue. To my understanding, when we talk about nH, we always refer to the absorption of the neutral gas (or dust grains) assuming a certain metallicity (as in the commonly-used absorption models, e.g., phabs, wabs, and tbabs). I'm not sure whether nH can be similarly defined or determined for (highly-)ionized gas, for example, in our case for SN 1987A.}

Located in the Large Magellanic Cloud (LMC), SN 1987A is the nearest supernova (SN) observed since Kepler's SN of 1604. {It was the Type II explosion of a blue supergiant. The peculiar CSM system of SN 1987A consists of three coaxial rings --- one equatorial ring (ER) and two side rings constituting an hourglass-like structure.
The SN blast wave encountered the innermost layer of the ER at $\sim4000$ days after the explosion, as indicated by the brightening of several ``hot spots'' in optical band and the corresponding excess of the soft X-ray flux \citep[e.g.,][]{1998ApJ...492L.139S,2000ApJ...537L.123L,2000ApJ...543L.149B,2002ApJ...567..314P}. Soon after, the front shock reached the main body of the ER at $\sim6000$ days, and has been heavily interacting with it since then \citep[e.g.,][]{2005ApJ...634L..73P,2009ApJ...703.1752R,2013ApJ...764...11H,2016ApJ...829...40F}, making SN 1987A one of the youngest and brightest SNRs.
The X-ray emission of SN 1987A is now dominated by the shock-heated plasma within the ER, including the dense clumps and the inter-clump materials, with a density $\sim10^3$--$10^4$\,cm$^{-3}$. On the other hand, the shocked low-density H II region ($\sim10^2$\,cm$^{-3}$), high-latitude materials beyond the ER, and the outermost-layer ejecta also make significant contributions \citep[e.g.,][]{2012ApJ...752..103D,2020A&A...636A..22O,2021ApJ...916...41S,2024ApJ...966..147R,2025inprep}.}
The small angular size \citep[$\sim2''$, e.g.,][]{2009ApJ...703.1752R} and high X-ray brightness \citep[$\sim7.8\times10^{-12}$\,erg\,cm$^{-2}$\,s$^{-1}$ in 0.5--2.0\,keV, e.g.,][]{2021ApJ...916...41S} of SN 1987A enable high-resolution X-ray spectroscopic studies using X-ray grating spectrometers, which have yielded fruitful results regarding the structure and evolution of the remnant, physical property of the shocked ejecta and CSM, collisionless shock heating mechanism, and so on \citep[e.g.,][]{2009ApJ...692.1190Z,2010A&A...515A...5S,2012ApJ...752..103D,2019NatAs...3..236M,2021ApJ...916...41S}. This also makes SN 1987A an excellent target in {the studies} of CX, RS, and hot gas absorption in SNRs.

{Such kind of studies relies on a comprehensive understanding of the physical properties of the {thermally}-emitting gas. 
Due to the complex structure of the remnant, the shock-heated plasmas in SN 1987A span a wide range of physical conditions (e.g., temperatures and ionization parameters), which has been extensively revealed by previous studies via X-ray spectral analysis \citep[e.g.,][]{2006A&A...460..811H,2004ApJ...610..275P,2006ApJ...646.1001P,2006ApJ...645..293Z,2009ApJ...692.1190Z,2010MNRAS.407.1157Z,2008ApJ...676..361H,2010A&A...515A...5S,2012ApJ...752..103D,2016ApJ...829...40F,2020ApJ...899...21B,2021ApJ...916...76A,2021ApJ...916...41S,2021ApJ...922..140R,2024ApJ...966..147R,2021ApJ...908L..45G,2022ApJ...931..132G,2022A&A...661A..30M}. As summarized by \citet[][see Table 6 therein]{2021ApJ...916...76A}, these works found plasma temperatures in the range of $\sim0.5$--$4$\,keV, with a low-temperature component at $\sim0.5$\,keV dominated by the dense ER, a high-temperature component at $\sim2$--$4$\,keV dominated by the H II region and the outermost ejecta, and some of them a third intermediate-temperature component at $\sim1$\,keV. The ionization parameters of the plasmas, on the other hand, also span a wide range from $\sim10^{10}$--$10^{13}$\,cm$^{-3}$\,s.}
Recently, \citet{2025inprep} obtained the continuous temperature distributions of SN 1987A by fitting the RGS plus EPIC-pn spectra with a differential emission measure (DEM) model. The obtained temperature distributions {reveal a major peak at $\sim0.6$--$1$\,keV with a high-temperature tail extending to $\gtrsim5$\,keV and} are well consistent with the {magnetohydrodynamic} simulations by \citet{2020A&A...636A..22O}, which helps to reveal the fading of the equatorial ring (ER) and the brightening of the shocked ejecta. Their results provided one of the most {accurate} depictions of the X-ray emission of SN 1987A, but still far from perfect. Since the DEM model considered only the thermal emission from the non-equilibrium ionization (NEI) plasmas, it is still worthwhile to take a further look into the residuals that may provide information about other mechanisms such as CX, RS, and hot gas absorption.

In this work, we extend the discussions in \citet{2025inprep} by examining the residuals left by the DEM modeling at O lines. We describe the observations and the reduction procedures in Section \ref{sec:obs}, revise and update the O line fluxes and line ratios in Section \ref{sec:result}, explore the physical origin of the unusual O line ratios in Section \ref{sec:discussion}, and finally make our conclusions in Section \ref{sec:conclusion}.

\section{Observations and Data Reduction}\label{sec:obs}

{We utilized {the same dataset adopted} in \citet{2025inprep}, which contains {XMM-Newton observations collected in 14 epochs} from 2007 to 2021.}
%SN 1987A has been regularly monitored in X-rays. In this work, we utilized a set of XMM-Newton observations taken since 2007, which is similar to those used in \citet{2021ApJ...916...41S}, but with two additions. One of the additional observation was taken in November 2020 (PI: F.~Haberl). And we obtained a new observation in December 2021 (PI: L.~Sun), representing the most recent evolution of SN 1987A. For each observation, we used both the RGS and the EPIC-pn exposures. The high energy resolution of RGS can help to detect and resolve the numerous emission lines in $0.35$--$2.5$\,keV, which is crucial for our spectral modeling. EPIC-pn has a large effective area and a complete energy coverage up to 10\,keV band, providing further constraints on the continuum and the high-temperature plasmas. The details of all observations are summarized in Table \ref{tab:obs}.
All the data were processed using the XMM-Newton Science Analysis Software (SAS, version 18.0.0)\footnote{https://www.cosmos.esa.int/web/xmm-newton/sas} with the latest calibration files. We extracted the EPIC-pn and RGS spectra following \citet{2021ApJ...916...41S}. The spectra were then optimally rebinned, adopting the \citet{2016A&A...587A.151K} optimal binning scheme.
{Due to the small angular size, SN 1987A cannot be spatially resolved by XMM-Newton, and thus was treated as a point-like source. Thereby, the extracted spectra represent the integrated X-ray emission from the whole remnant, and the RGS spectral resolution ($\sim2$\,eV at $\sim0.6$\,keV for O lines) will not be affected by the source extent.}

{Unless otherwise specified, in this paper the metal abundances are presented with respect to their solar values \citep{2000ApJ...542..914W}, and the error bars represent the 1-$\sigma$ uncertainties.}

%%\begin{deluxetable*}{ccccccc}
%%	\tablecaption{Observations\label{tab:obs}}
%%	\tablenum{1}
%%	\tablehead{
%%		\multicolumn{3}{c}{}&\multicolumn{2}{c}{$t_{\rm exp}$ (ks)\tablenotemark{a}}&\multicolumn{2}{c}{$\sum$GTI (ks)\tablenotemark{b}}\\
%%		ObsID & Date&Age (days) & EPIC-pn & RGS & EPIC-pn & RGS
%%        }
%%	\startdata
%%	    0406840301 & 2007 Jan 17 & 7267 & 106.9 & 111.3 & 61.1 & 109.8 \\
%%	    0506220101 & 2008 Jan 11 & 7627 & 110.1 & 114.3 & 70.7 & 102.4 \\
%%		0556350101 & 2009 Jan 30 & 8012 & 100.0 & 101.9 & 66.4 & 101.8 \\
%%		0601200101 & 2009 Dec 11 & 8327 & 89.9 & 91.8 & 82.4 & 91.7 \\
%%		0650420101 & 2010 Dec 12 & 8693 & 64.0 & 65.9 & 52.7 & 65.9 \\
%%		0671080101 & 2011 Dec 2 & 9048 & 80.6 & 82.5 & 64.2 & 80.6 \\
%%		0690510101 & 2012 Dec 11 & 9423 & 68.0 & 69.9 & 59.4 & 69.8 \\
%%		0743790101 & 2014 Nov 29 & 10141 & 78.0 & 79.6 & 56.4 & 79.4 \\
%%		0763620101 & 2015 Nov 15 & 10492 & 64.0 & 65.9 & 58.0 & 65.8 \\
%%		0783250201 & 2016 Nov 2 & 10845 & 72.4 & 74.3 & 50.3 & 74.2 \\
%%		0804980201 & 2017 Oct 15 & 11192 & 77.5 & 79.4 & 27.8 & 79.3 \\
%%		0831810101 & 2019 Nov 27 & 11964 & 32.4 & 34.9 & 11.1 & 34.8 \\
%%      0862920201 & 2020 Nov 24 & 12328 & 77.6 & 79.7 & 57.8 & 72.8 \\
%%        0884210101 & 2021 Dec 28 & 12727 & 90.6 & 89.4 & 75.2 & 86.1 \\
%%	\enddata
%%	\tablenotetext{a}{Total exposure times.}
%%	\tablenotetext{b}{Total good time intervals after background flare removal.}
%%\end{deluxetable*}

\section{{Oxygen line ratios}}\label{sec:result}

\begin{figure*}[ht]
    \centering
    \includegraphics[width=\linewidth]{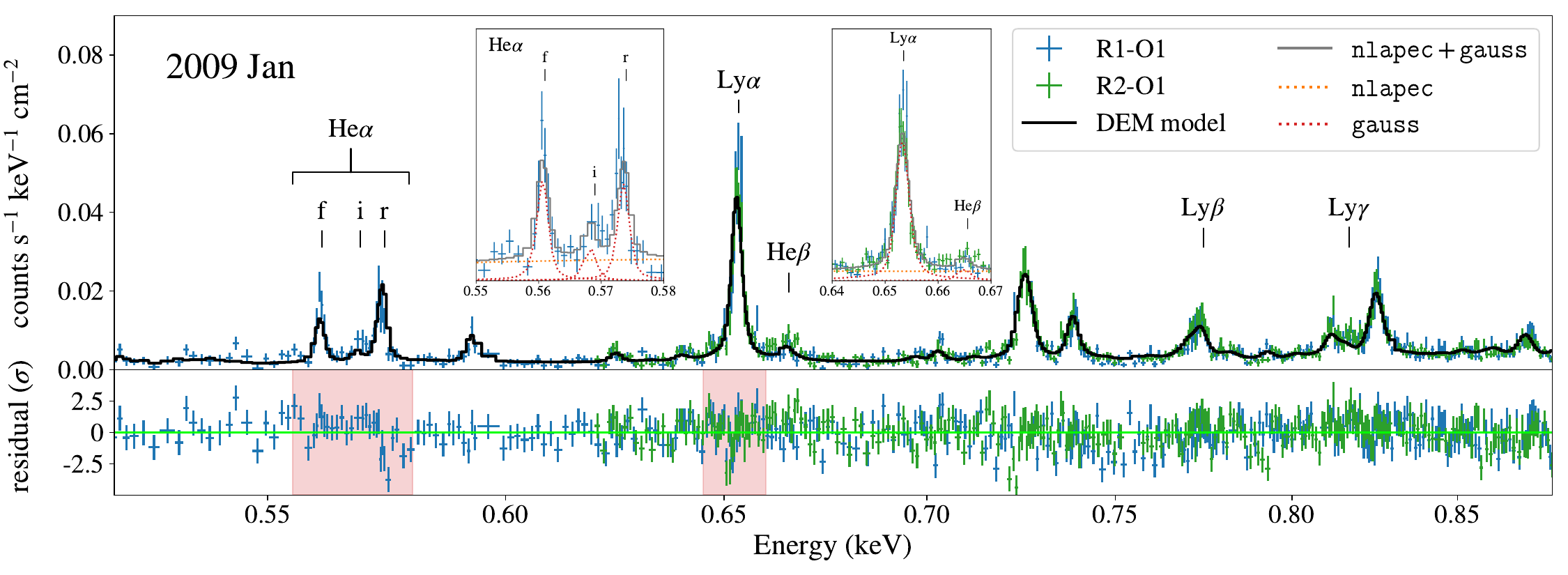}
    \caption{{An example of the O-emitting band spectra of SN 1987A fitted with the DEM model \citep{2025inprep}. The major O lines are labeled. The red boxes highlight the major residuals left in the DEM modeling, where the model overestimated the O VII resonance line and the O VIII Ly$\alpha$ lines while underestimated the O VII forbidden line. The two subplots show the zoom-in view of O VII He$\alpha$ and O VIII Ly$\alpha$, fitted with the ``{\tt nlapec} $+$ {\tt gauss}'' model.}}
    \label{fig:O_line_spec}
\end{figure*}

\begin{figure*}[ht]
    \centering
    \includegraphics[width=\linewidth]{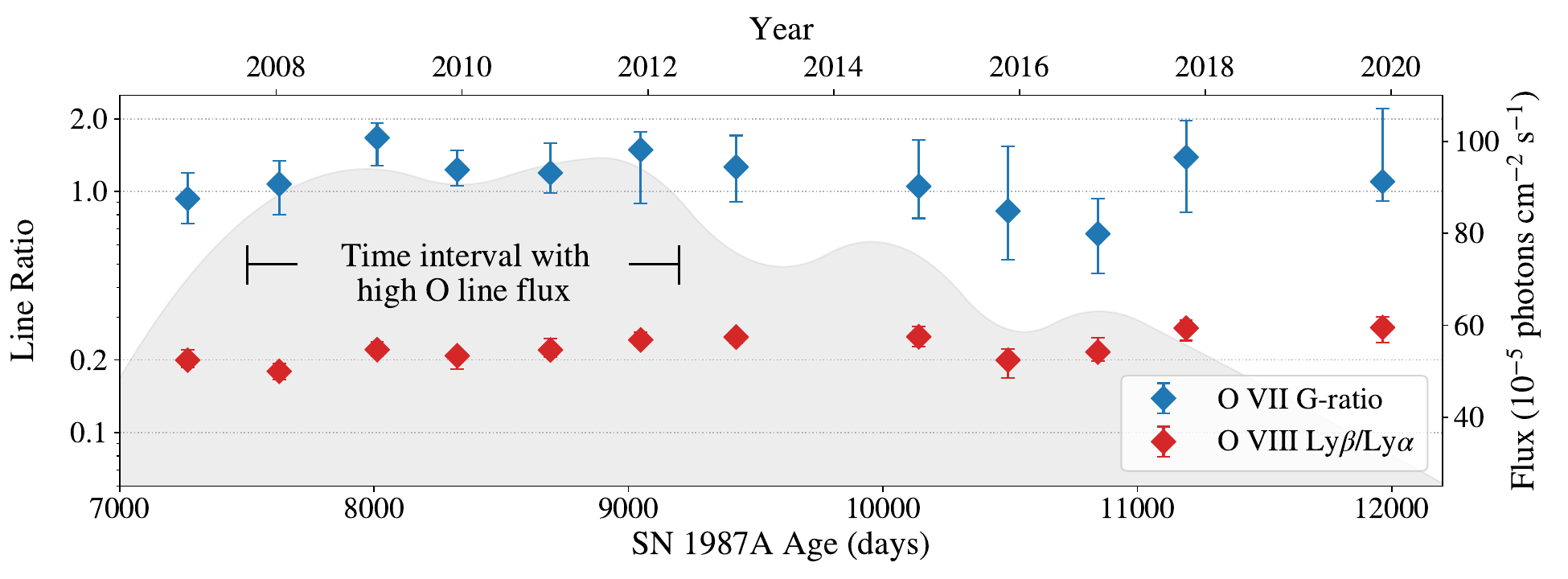}
    \caption{O line ratios and their temporal evolutions in SN 1987A. The blue and red data points denote the O VII G-ratio and the O VIII Ly$\beta$/Ly$\alpha$ flux ratio, respectively. The gray under-filled curve indicates the temporal evolution of the total O line flux (O VII He$\alpha$ $+$ O VIII Ly$\alpha$).}
    \label{fig:line_ratio_obs}
\end{figure*}

{The X-ray spectra of SN 1987A have been extensively analyzed based on different observations with different spectral models \citep[e.g.,][]{2009ApJ...692.1190Z,2010A&A...514L...2M,2012ApJ...752..103D,2021ApJ...916...41S,2021ApJ...916...76A,2021ApJ...922..140R,2022A&A...661A..30M}. Recently, \citet{2025inprep} successfully fitted the X-ray spectra of SN 1987A using a DEM model and provided significantly improved results (fit statistics) compared with the previous discrete models \citep[e.g., a 2-T model as in][]{2021ApJ...916...41S}. 
{With the DEM model, \citet{2025inprep} characterized the continuous distribution of the X-ray gas in SN 1987A in a temperature range of $0.1$--$10$\,keV. They assumed a powerlaw relation between the ionization parameter and the plasma temperature, and an identical metal abundance among different temperature components. Their results revealed a major peak in the DEM distribution at $\sim0.5$--$1$\,keV, with a high-temperature tail extend to $\gtrsim5$\,keV.}
Here, we adopted their best-fit models and compared them with the RGS spectra of SN 1987A. 
We found the DEM models still leave some residuals in the fitted spectra.}
%The DEM model is successful in fitting the X-ray spectra of SN 1987A and provides significantly improved results (fit statistics) compared with the discrete models. However, we still found some residuals in the fitted spectra. 
The most prominent residual features are lying in the O VII triplets and the O VIII Ly$\alpha$ lines, where the model overestimated the O VII resonance line and the O VIII Ly$\alpha$ lines while underestimating the O VII forbidden line (Figure \ref{fig:O_line_spec}). %\lsun{response to SO ``{\it - SO: Looking at Fig. 1, I see some other important residuals (for instance in the [0.7, 0.75] keV band). Why not focusing also on the origin of these residuals?}'': Yes, there are some other residuals. For example, some residuals can be seen at $\sim0.725$\,keV, as you mentioned, which may be related to Fe XVII 17\,${\rm \AA}$ resonance line; some residuals seen at $\sim0.825$\,keV may be related to Fe XVII 15\,${\rm \AA}$ resonance line; and some residuals appear at O VIII Ly$\beta$, Ly$\gamma$ lines. However, these residuals can only be seen in some of the observations, and are less significant compared to those at O VII triplet and O VIII Ly$\alpha$. As a test, I have tried to add Gaussian absorption components at Fe XVII 15\,${\rm \AA}$/17\,${\rm \AA}$ lines. I obtained no significant detection of the absorption features, with only upper limits may be estimated. On the other hand, the residuals at O VIII Ly$\beta$/Ly$\gamma$ lines seems to be due to the underestimation of the O abundance, since they are reduced after adding the Gaussian absorption components at O VII resonance line and O VIII Ly$\alpha$ line.}

\begin{deluxetable*}{cccccccccc}
    \tablecaption{Oxygen Line Fluxes and Ratios of SN 1987A\label{tab:O_line}}
    \tablenum{1}
    \tablehead{
    \multirow{3}{*}{Obs. Date} & \multirow{3}{*}{Age (days)} &
    \multicolumn{5}{c}{Line Flux ($10^{-5}$\,photon\,cm$^{-2}$\,s$^{-1}$)} & & \multicolumn{2}{c}{Line Ratio} \\
    \cline{3-7}\cline{9-10}
     & & O VII f & O VII i & O VII r & O VIII Ly$\alpha$ & O VIII Ly$\beta$ & & \multirow{2}{*}{G-ratio} & \multirow{2}{*}{Ly$\beta$/Ly$\alpha$}
     \\
     & & {(0.5611\,keV)} & {(0.5686\,keV)} & {(0.5740\,keV)} & {(0.6537\,keV)} & {(0.7746\,keV)} & & & 
    }
    \startdata
	    2007 Jan 17 & 7267 & $12.64^{+1.51}_{-2.18}$ & $1.26^{+3.05}_{-0.93}$ & $14.91^{+1.87}_{-1.89}$ & $34.47^{+1.03}_{-0.80}$ & $6.88^{+0.68}_{-0.42}$ && $0.93^{+0.26}_{-0.20}$ & $0.20^{+0.02}_{-0.01}$ \\
	    2008 Jan 11 & 7627 & $15.06^{+2.35}_{-2.89}$ & $4.28^{+2.45}_{-1.68}$ & $18.03^{+3.41}_{-3.19}$ & $43.24^{+2.20}_{-2.47}$ & $7.76^{+0.37}_{-0.43}$ && $1.07\pm0.27$ & $0.18\pm0.01$ \\
		2009 Jan 30 & 8012 & $18.07^{+1.08}_{-1.41}$ & $5.81^{+1.94}_{-2.89}$ & $14.32^{+2.69}_{-1.72}$ & $45.63^{+0.81}_{-1.85}$ & $10.07^{+0.63}_{-0.56}$ && $1.67^{+0.25}_{-0.39}$ & $0.22^{+0.02}_{-0.01}$ \\
		2009 Dec 11 & 8327 & $13.65^{+2.08}_{-1.73}$ & $3.66^{+1.14}_{-0.59}$ & $14.09^{+1.36}_{-2.13}$ & $49.01^{+2.90}_{-1.37}$ & $10.20^{+0.42}_{-1.07}$ && $1.23^{+0.25}_{-0.18}$ & $0.21^{+0.01}_{-0.03}$ \\
		2010 Dec 12 & 8693 & $16.69^{+3.02}_{-0.99}$ & $0.32^{+2.69}_{-0.24}$ & $14.26^{+2.30}_{-3.18}$ & $52.22^{+0.50}_{-1.48}$ & $11.49^{+1.28}_{-0.76}$ && $1.19^{+0.39}_{-0.21}$ & $0.22^{+0.03}_{-0.01}$ \\
		2011 Dec 2 & 9048 & $14.08^{+2.31}_{-2.92}$ & $4.23^{+1.15}_{-2.62}$ & $12.32^{+4.16}_{-1.47}$ & $51.07^{+1.37}_{-2.27}$ & $12.37^{+0.78}_{-0.57}$ && $1.49^{+0.27}_{-0.59}$ & $0.24^{+0.02}_{-0.01}$ \\
		2012 Dec 11 & 9423 & $14.00^{+2.79}_{-1.22}$ & $<3.28$ & $11.09^{+3.00}_{-1.86}$ & $48.57^{+1.15}_{-0.41}$ & $12.09^{+0.58}_{-0.69}$ && $1.26^{+0.44}_{-0.36}$ & $0.25^{+0.01}_{-0.02}$ \\
		2014 Nov 29 & 10141 & $7.61^{+2.02}_{-1.81}$ & $6.95^{+2.48}_{-1.55}$ & $13.88^{+2.86}_{-3.37}$ & $37.28^{+0.78}_{-1.02}$ & $9.31^{+0.92}_{-0.83}$ && $1.05^{+0.58}_{-0.28}$ & $0.25^{+0.03}_{-0.02}$ \\
		2015 Nov 15 & 10492 & $1.75^{+2.70}_{-1.31}$ & $5.60^{+2.85}_{-1.90}$ & $8.87^{+1.73}_{-1.05}$ & $35.75^{+1.35}_{-1.18}$ & $7.14^{+0.78}_{-1.07}$ && $0.83^{+0.71}_{-0.31}$ & $0.20^{+0.02}_{-0.03}$ \\
		2016 Nov 2 & 10845 & $5.69^{+1.83}_{-1.32}$ & $1.42^{+1.50}_{-1.07}$ & $10.66^{+2.15}_{-2.39}$ & $37.22^{+0.92}_{-1.47}$ & $8.03^{+1.14}_{-0.66}$ && $0.67^{+0.26}_{-0.21}$ & $0.22^{+0.03}_{-0.02}$ \\
		2017 Oct 15 & 11192 & $7.57^{+2.50}_{-2.56}$ & $3.41^{+1.60}_{-2.16}$ & $7.93^{+2.16}_{-1.32}$ & $28.93^{+1.47}_{-1.39}$ & $7.84^{+0.52}_{-0.78}$ && $1.38^{+0.58}_{-0.57}$ & $0.27^{+0.02}_{-0.03}$ \\
		2019 Nov 27 & 11964 & $<0.22$ & $5.19^{+1.66}_{-0.76}$ & $4.73^{+0.42}_{-0.44}$ & $25.87^{+0.30}_{-0.24}$ & $7.05^{+0.73}_{-0.94}$ && $1.10^{+1.10}_{-0.19}$ & $0.27^{+0.03}_{-0.04}$ \\
        2020 Nov 24 & 12328 & $<2.86$ & $5.62^{+3.41}_{-3.23}$ & $7.48^{+3.28}_{-3.12}$ & $30.71^{+1.82}_{-1.69}$ & $6.55^{+1.32}_{-1.38}$ && $0.79^{+0.66}_{-0.55}$ & $0.21^{+0.04}_{-0.05}$ \\
        2021 Dec 28 & 12727 & $3.49^{+1.96}_{-2.01}$ & $<1.48$ & $<3.33$ & $22.68\pm1.51$ & $4.22^{+0.90}_{-0.87}$ && --- & $0.19\pm0.04$ \\
    \enddata
    %\tablecomments{}
\end{deluxetable*}

{\citet{2021ApJ...916...41S} measured the fluxes of 36 emission lines in the X-ray spectrum of SN 1987A and followed their evolution from 2007 to 2019 by fitting the RGS spectra with a model containing one continuum {(described by the {\tt nlapec}\footnote{https://heasarc.gsfc.nasa.gov/xanadu/xspec/manual/XSmodelNlapec.html} model which includes the thermal bremsstrahlung, the radiative recombination continua, and the two-photon emission)} plus multiple gaussian lines {(described by several {\tt Gauss} components)}. Here, we revised and updated their results by applying the same procedure to the updated dataset which contains two additional observations taken in 2020 and 2021.
We measured the fluxes of the most prominent O lines (i.e., the forbidden, intercombination, and resonance line for O VII, and the Ly$\alpha$ and Ly$\beta$ lines for O VIII). 
Based on these updated line fluxes, we calculated the O VII G-ratio (defined as $\mathcal{G}\equiv(F+I)/R$, where $F$, $I$, {and} $R$ stand for the flux of the forbidden, the intercombination, and the resonance line, respectively) and the O VIII Ly$\beta$/Ly$\alpha$ ratio for SN 1987A from 2007 to 2021. The line fluxes and ratios are summarized in Table \ref{tab:O_line}. {Given the complex structure of SN 1987A (as described in Section \ref{sec:intro}), one should keep in mind that different components of the remnant, such as the shocked ER, H II region, high-latitude materials, and outer-layer ejecta, contribute to the O lines simultaneously. Thereby, the O line fluxes and ratios measured here represents the average properties of all these components with a large range of physical conditions.} 

{For collisional ionization equilibrium (CIE) or NEI (under-ionized) plasma, the O VII G-ratio is commonly in the range of $\sim0.5$--$1$ varying with temperature and ionization parameter, while O VIII Ly$\beta$/Ly$\alpha$ commonly in the range of $\sim0$--$0.15$ varying with temperature \citep[e.g.,][and see the discussions in Section \ref{sec:nei}]{1999LNP...520..109M,2008SSRv..134..155K,2010SSRv..157..103P}. However, according to the above results,} we found a high O VII G-ratio $\gtrsim1$ and a high O VIII Ly$\beta$ to Ly$\alpha$ ratio $\gtrsim0.2$ in SN 1987A, especially in the epoch from 2009 to 2012} when the O lines were around their maximum luminosity (Figure \ref{fig:line_ratio_obs}). 

We note that the measured O VIII Ly$\beta$ flux may be affected by Fe L lines, such as the contribution from the Fe XVIII F6 line \citep[$2s^2 2p^4 3s~^2{\rm P}_{3/2} \rightarrow 2s^2 2p^5~^2{\rm P}_{3/2}$ at 0.775\,keV, line label adopted from][]{1998ApJ...502.1015B}, which may lead to an overestimated O VIII Ly$\beta$/Ly$\alpha$ ratio.
{The soft-band X-ray emission in SN 1987A is dominated by a low-temperature plasma component with $kT_{\rm e}\sim0.6$\,keV and $n_{\rm e}t\sim10^{11}$\,cm$^{-3}$\,s according to previous studies \citep[e.g.,][]{2010A&A...515A...5S,2021ApJ...916...76A,2021ApJ...922..140R,2022A&A...661A..30M}. Under this condition, we calculated the emissivities of O VIII Ly$\beta$ and Fe XVIII F6, and found the relative contribution of Fe XVIII F6 line ($\epsilon_{\rm F6}/(\epsilon_{\rm F6}+\epsilon_{\rm Ly\beta})$) to be $\sim17\%$. Nevertheless, the O and Fe L emission could be contributed by plasma that spans a larger range of temperatures and ionization parameters as shown in \citet{2025inprep}. We then explored a larger parameter space, where temperature varies between 0.3\,keV to 3\,keV and ionization parameter varies between $10^{11}$\,cm$^{-3}$\,s to $10^{12}$\,cm$^{-3}$\,s. The obtained emissivity-weighted average contribution of Fe XVIII F6 line is $\sim25\%$. As a result, we consider the O VIII Ly$\beta$/Ly$\alpha$ ratio measured above might be overestimated by a factor $\sim0.3$ due to the potential contribution of Fe XVIII F6. However, the O VIII Ly$\beta$/Ly$\alpha$ even after correcting this effect will still be $\gtrsim0.15$, which is difficult to fully explained with the common CIE/NEI plasma models (see the discussions in Section \ref{sec:nei}).}

\section{{Physical origin of the unusual O line ratios}}\label{sec:discussion}

%High O VII G-ratio (or high forbidden-to-resonance line ratio) has been found in several SNRs, such like Cygnus Loop \citep{2011ApJ...730...24K,2015MNRAS.449.1340R,2019ApJ...871..234U}, Puppis A \citep{2012ApJ...756...49K}, N49 \citep{2020ApJ...897...12A}, N132D \citep{2020ApJ...900...39S}, G296.1$-$0.5 \citep{2022ApJ...933..101T}, J0453.6$-$6829 \citep{2022PASJ...74..757K}, etc. The physical origin of the enhanced line ratios is still under debate, while the major possible mechanisms include the NEI effects, the charge exchange (CX) processes, the resonant scattering (RS), and the absorption of foreground hot gas. 
{As mentioned in the introduction, a high O VII G-ratio has been observed in several SNRs. Despite efforts to understand its origin, it remains unclear whether the primary mechanism responsible is NEI effects, CX processes, RS, or absorption by foreground hot gas.}
We then discuss these possibilities as below.

    \subsection{NEI effects} \label{sec:nei}

    \begin{figure*}[ht]
    \centering
    \includegraphics[width=\linewidth]{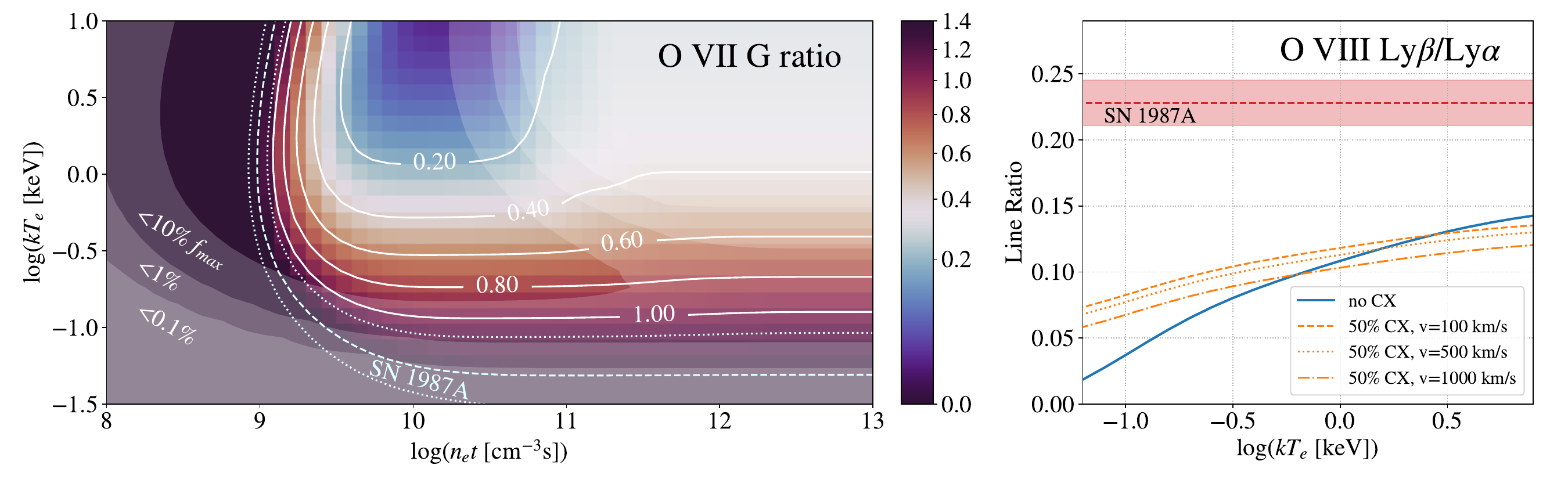}
    \caption{Left: O VII G-ratio as a function of $kT_{\rm e}$ and $n_{\rm e}t$ for ionizing NEI plasma. The observed G-ratio for SN 1987A is indicated by the cyan contours, {where the dashed line denotes the average value and dotted lines indicate the error range}. The shaded contour regions indicate the levels of total O VII He$\alpha$ flux, as $<10\%$, $<1\%$, and $<0.1\%$ of the maximum flux, respectively. Right: O VIII Ly$\beta$/$\alpha$ ratio as a function of $kT_{\rm e}$, for both NEI and CIE plasma (NEI effects make little impact on the Ly$\beta$/Ly$\alpha$ ratio, which is similar to that in CIE scenario). {The blue solid line denotes no CX contribution, while the dashed, dotted, and dash-dotted orange lines denoting $50\%$ CX contribution with collision velocity of 100\,km\,s$^{-1}$, 500\,km\,s$^{-1}$, and 1000\,km\,s$^{-1}$, respectively.} The observed flux ratio for SN 1987A is indicated by the red region.}
    \label{fig:line_ratio_compared}
    \end{figure*}
    
    The NEI effects of plasmas may either enhance (e.g., at the early stage of the ionization when the inner-shell ionization of Li-like ions is boosted due to a large population of Li-like ions, which enhances the forbidden lines) or suppress (e.g., during the ionizing process when the populations of Li-like and H-like ions are both low, and thus the inner-shell ionization and the recombination processes being quenched, which suppresses the forbidden lines) the G-ratio of He-like ions compared with that in CIE scenario \citep[see, e.g.,][for a detailed review]{2010SSRv..157..103P}. For SN 1987A, it is clear that the majority of the X-ray plasma is in the NEI state, i.e., still ionizing \citep[e.g.,][]{2010A&A...515A...5S,2021ApJ...916...41S}. {However, even though the previous studies such as \citet{2025inprep} have been using the NEI plasma models for spectral fitting, it is still possible that these models fail to reproduce the line ratios due to the limited model setup and simplified parameter configurations.} Therefore, we firstly explored the possibility of reproducing the observed O VII G-ratios within the NEI regime. 
    
    We evaluated the O VII G-ratio in a large $kT_{\rm e}$-$n_{\rm e}t$ space, as illustrated by Figure \ref{fig:line_ratio_compared} (the calculations were performed using SPEX version 3.07.03\footnote{https://www.sron.nl/astrophysics-spex/}). We found the observed O VII G-ratio in SN 1987A can only be reproduced in extreme conditions, i.e., at a very low temperature ($kT_{\rm e}\lesssim0.1$\,keV) or at the very early ionization stage ($n_{\rm e}t\sim10^9$\,cm$^{-3}$\,s). However, at a very low temperature, even though the high G-ratio can be achieved, the O VII lines are rather faint ($<1\%$ of their maximum flux). Thereby the contributions of these low-temperature plasmas can be ignored. On the other hand, since the average density of SN 1987A is pretty high \citep[$\gtrsim10^3$\,cm$^{-3}$, e.g.,][]{2021ApJ...916...41S}, the plasma will quickly go through the $n_{\rm e}t\sim10^9$\,cm$^{-3}$\,s stage, and move into $n_{\rm e}t\sim10^{10-12}$\,cm$^{-3}$\,s. Thus the contributions of these low-$n_{\rm e}t$ plasmas are also negligible.
    
    In a word, the observed O VII G-ratio cannot be fully explained under the NEI regime. This can be further demonstrated by taking other line ratios into consideration. As shown in Figure \ref{fig:line_ratio_diagnostic}, by plotting the observed Ly$\alpha$/He$\alpha$ ratio together with the G-ratio in the $kT_{\rm e}$-$n_{\rm e}t$ diagram, we found that the two line ratios are not consistent with each other --- there is no [$kT_{\rm e}$, $n_{\rm e}t$] combination which can reproduce both the O VII G-ratio and the Ly$\alpha$/He$\alpha$ ratio simultaneously. {However, this inconsistency is only seen for O, while for Ne and Mg the two ratios provide reasonable estimations for plasma temperature and ionization parameter as $kT_{\rm e}\sim0.4$--$1$\,keV and $n_{\rm e}t\sim10^{11}$--$10^{12}$\,cm$^{-3}$\,s. These are comparable with previous results, such like the low-temperature ($\sim0.4$\,keV) and middle-temperature ($\sim0.8$\,keV) components of the 3-T {\tt vnei} modeling in \citet{2022ApJ...931..132G}, representing the X-ray emission dominated by the shocked ER.}

    Over-ionized (recombining) plasma may also lead to enhanced G-ratios. However, this is unlikely to be the case for SN 1987A, since no other recombination features (such as radiative recombination {continua}) have been observed.

    NEI effects make little impact on the Ly$\beta$/Ly$\alpha$ ratio, which is similar to that in CIE scenario as a function of temperature (Figure \ref{fig:line_ratio_compared}). Therefore, the high O Ly$\beta$/Ly$\alpha$ ratio cannot be explained in NEI regime either.

    \begin{figure}[ht]
    \centering
    \includegraphics[width=\linewidth]{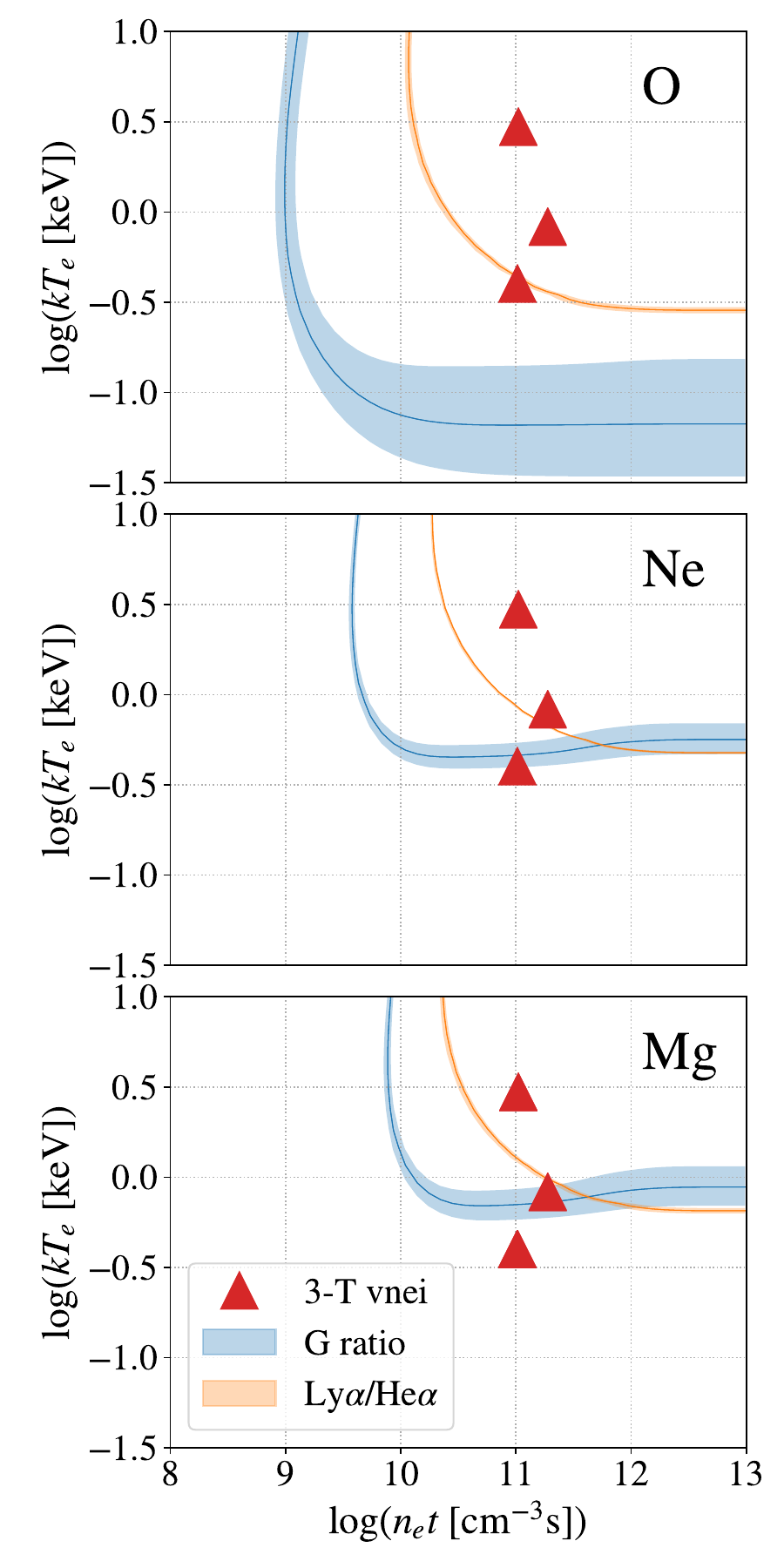}
    \caption{$kT_{\rm e}$ and $n_{\rm e}t$ parameter spaces confined by the observed G-ratios and Ly$\alpha$/He$\alpha$ ratios of O, Ne, and Mg for SN 1987A. 3-T {\tt vnei} results are adopted from \citet{2022ApJ...931..132G}}
    \label{fig:line_ratio_diagnostic}
    \end{figure}
    
    \subsection{Charge exchange}

    The {CX} process involves an electron transferring from one atom or ion to another. Specifically, the CX relevant to X-ray astrophysics usually refers to a donor H/He atom colliding with a highly-ionized ion (e.g., O$^{+7}$, O$^{+8}$). During the collision, an electron may be transferred from the atom to the ion, recombined to an highly-excited state, which will then cascade to the ground level and result in a series of emission lines in the X-ray spectrum. In the case of charge exchanged to a H-like O ion (${\rm H}+{\rm O}^{+7}\rightarrow {\rm H}^{+}+{\rm O}^{+6}$), the emission of the recombined, excited O$^{+6}$ ion is dominated by a bright O VII forbidden line, which will lead to a large G-ratio \citep[e.g.,][]{2012AN....333..301S,2016A&A...588A..52G,2023arXiv230111335G}.

    At the outermost layer of SNR, the upstream neutral atoms will pass through the shock front and interact with the downstream hot ions, which provides a promising site for CX to take place. Observational evidence of CX emission has been found in the optical band for many SNRs as the broad-line component in Balmer-dominated (non-radiative) shocks \citep[e.g., Tycho, SN 1006, Cygnus Loop, RCW 86, SNR 0509-67.5,][]{1987ApJ...315L.135K,2001ApJ...547..995G,2010ApJ...719L.140H}. However, in X-rays there are only a few cases that have been indicated \citep[e.g., Cygnus Loop, Puppis A, G296.1$-$0.5,][]{2011ApJ...730...24K,2015MNRAS.449.1340R,2012ApJ...756...49K,2022ApJ...933..101T}. 
    {Evidence of X-ray CX emission has been found in other (but relevant) astrophysical environments, such like at the interface between old overlapping SNRs and molecular clouds \citep[e.g., in the Carina and 30 Doradus star forming complexes,][]{2011ApJS..194...15T,2011ApJS..194...16T,2024ApJS..273....5T}}. 
    We then explored whether CX may help to explain the observed line ratios in SN 1987A.

    {We added a CX component (described by the {\tt vacx2} model in PyXSPEC) to the DEM model adopted in \citet{2025inprep}, {in order to see whether it could improve the spectral fits.}} We fixed the collision velocity as 500\,km\,s$^{-1}$, while letting the plasma temperature and normalization free to vary. We fitted the SN 1987A spectra in all epochs with this model including CX emission. However, we only found considerable improvement in the C statistics in one observation (ObsID 0556350101 taken in Jan 2009), for which we got $\Delta C=-13.5$ with $\Delta{\rm d.o.f.}=-2$. For other observations, this model provides no significant improvements to the fitting ($\overline{\Delta C}\sim-1.8$ with $\Delta{\rm d.o.f.}=-2$). {The CX emission may arise from plasmas with different collision velocities and temperatures. We have investigated into this effect by changing the collision velocity from 50\,km\,s$^{-1}$ to 1000\,km\,s$^{-1}$ and by adding another {\tt vacx2} component with different temperature, while the resulting C-stat show little changes.}

    The major problem of the CX scenario is that it can not explain the enhanced Ly$\beta$/Ly$\alpha$ ratio, or say the suppressed Ly$\alpha$ flux. On the contrary, the O VIII Ly$\alpha$ line should also be boosted by CX, {as long as} there are enough O$^{+8}$ ions.
    {In order to quantify the effects of CX emission on Ly$\beta$/Ly$\alpha$, we calculated the O VIII Ly$\alpha$ and Ly$\beta$ fluxes and derived the line ratios under the CX scenario. We explored a large parameter space, where the plasma temperature varies in the range of 0.1--10\,keV and the collision velocity varies in the range of 50--1000\,km\,s$^{-1}$. We found that the O VIII Ly$\beta$/Ly$\alpha$ ratio mainly depends on the collision velocity, where the line ratio decreases from $\sim0.128$ to $\sim0.098$ with the collision velocity increasing from 50\,km\,s$^{-1}$ to 1000\,km\,s$^{-1}$. Given a certain collision velocity, the line ratio shows no significant variation with the varying temperature. Assuming a contribution of CX emission as high as $50\%$ to the total O VIII emission, it will slightly increase the line ratio at low temperatures while slightly decrease the line ratio at high temperatures (as seen in the right panel of Figure \ref{fig:line_ratio_compared}). However, in general the expected Ly$\beta$/Ly$\alpha$ ratio is still $\lesssim0.15$, and thus cannot explain the observations.}
    
    \subsection{Resonant scattering}\label{sec:rs}

    {In modeling the X-ray emission of} SNRs, it is commonly assumed that the plasmas are optically thin. However, this may not be the case under certain conditions, especially for the emission from resonance lines. The photon emitted from a resonant (allowed) transition can be efficiently (characterized by a larger cross-section than forbidden transitions) absorbed by a suitable ion through resonant absorption. And for a resonant absorption from $n=1$ to $n=2$, the excited state will decay immediately by emitting a photon with roughly the same energy, but in a random direction. Therefore, when talking about the resonant absorption happening within the source, it is often referred to as ``resonant scattering'', {since there is usually no reduction in the number of resonance line photons.}
    %since the photon has not been that much destroyed. 
    If the optical depth is large enough (either due to a large length scale or a high density of the source), RS may become significant and alter the resonance line flux and its surface brightness distribution.

    In X-rays, RS effects have been mostly studied in diffuse hot plasma of elliptical galaxies, galaxy clusters, and galactic bulges \citep[e.g.,][]{2002ApJ...579..600X,2018PASJ...70...10H,2018ApJ...861..138C,2023arXiv231003892C}, for which the RS-modified line fluxes and {surface brightness} distributions have been used to infer the turbulent Mach numbers. As for SNRs, {there have been a few indications for its presence,}
    %there are only a few evidences have been indicated, 
    including the enhanced O VII G-ratios observed in DEM L71 \citep{2003A&A...406..141V}, Cygnus Loop \citep{2019ApJ...871..234U}, and N49 \citep{2020ApJ...897...12A}. Despite the small size of SN 1987A, its rather high density \citep[$\sim2400$\,cm$^{-3}$ for the shocked ER,][]{2021ApJ...916...41S} makes the RS effects still possible to be significant. Here we first make a rough estimation of the RS optical depths {of O VII resonance line and O VIII Ly$\alpha$ line} in SN 1987A {(RS makes little effect on O VII forbidden and intercombination lines and O VIII Ly$\beta$, given their small oscillator strengths, i.e., $\sim2.0\times10^{-10}$, $\sim8.2\times10^{-5}$, and $\sim3.9\times10^{-2}$, respectively)}.

    The optical depth of the resonant absorption (scattering) at the line centroid can be evaluated as \citep{1995A&A...302L..13K}:
    \begin{equation}\label{eq:1}
        \tau=\frac{4.24\times10^6fN_{\rm H,20}\left(\frac{n_{\rm i}}{n_{\rm Z}}\right)\left(\frac{n_{\rm z}}{n_{\rm H}}\right)\left(\frac{M}{T_{\rm keV}}\right)^{\frac{1}{2}}}{E_{\rm eV}\left(1+\frac{0.0522M\nu^2_{100}}{T_{\rm keV}}\right)^{\frac{1}{2}}}
    \end{equation}
    where $f$ is the oscillator strength of the line, $N_{\rm H,20}$ the hydrogen column density in unit of $10^{20}$\,cm$^{-2}$, $n_{\rm i}$ the ion density, $n_{\rm Z}$ the density of the element, $n_{\rm H}$ the hydrogen density, $M$ the atomic weight of the ion, $T_{\rm keV}$ the ion temperature in keV, $E_{\rm eV}$ the line centroid energy in eV, and $\nu_{100}$ the micro turbulence velocity in unit of 100\,km\,s$^{-1}$. Adopting a distance to SN 1987A of $51.4$\,kpc \citep{2005coex.conf..585P}, an angular size of $\sim2''$ \citep[e.g.,][]{2009ApJ...703.1752R}, a hydrogen density of $\sim2400$\,cm$^{-3}$ \citep{2021ApJ...916...41S}, an O abundance of $\sim0.18$ solar abundance \citep[e.g.,][]{2021ApJ...916...41S,2022ApJ...931..132G}, a turbulence velocity of $\sim500$\,km$^{-1}$ \citep[evaluated from the measured O VII line width by][]{2010A&A...515A...5S}, and an He-like O ion fraction of $\sim0.14$ \citep[taking $kT_{\rm e}=0.4$\,keV and $n_{\rm e}t=10^{11}$\,cm$^{-3}$\,s, e.g.,][]{2022ApJ...931..132G}, we got an estimation on the O VII resonance line optical depth as $\tau_{\rm OVII}\sim2.4$ {and O VIII Ly$\alpha$ optical depth as $\tau_{\rm OVIII}\sim2.2$.}\footnote{Here we assumed the O ion temperature is equal to the electron temperature. It is possible that the ion temperature is much higher than the electron/proton temperature \citep[e.g.,][]{2019NatAs...3..236M}. However, even if we take an O ion temperature as high as 20 times of electron temperature (i.e., $\sim8$\,keV), we still got an optical depth $\sim2$. And in that case, the inferred micro turbulence velocity will be much lower, since the observed line width will be more attributed by thermal broadening.}
    {Given the fact that the shocked gas in SN 1987A spans a large range of physical properties and considering the uncertainties in previous studies, we then estimated a range of the optical depths. For the plasma with temperature in the range of 0.3--1\,keV and ionization parameter in the range of 10$^{11}$--10$^{12}$\,cm$^{-3}$\,s (which may be the major plasma component that contribute to O line emission), we got the optical depth of O VII resonance line in the range of $\sim0.01$--$5$, with an emissivity-weighted average value of $\sim2.7$, and the optical depth of O VIII Ly$\alpha$ in the range of $\sim0.01$--$2.3$, with an emissivity-weighted average value of $\sim1.3$.}

    However, a large optical depth of RS does not necessarily lead to the change of observed resonance line flux. It highly depends on in what geometry the hot gas is distributed in the remnant and in what perspective we observe it. For example, if we observe the resonance line emission from the whole region of a spherically symmetric SNR, the RS will not make any difference on the integral line flux. However, the {surface brightness} distribution of the line may be modified, and the line flux from a certain sub-region of the remnant can be changed. As shown by \citet{2024ApJ...967...99L} using MC simulations, for a spherically symmetric SNR with a Sedov-Tylor gas distribution, if the optical depth is large enough, the resonance line flux will be significantly suppressed at the outer dense shell, but be enhanced at the inner low-density region of the remnant. On the other hand, an SNR with highly asymmetric geometry, just like SN 1987A, may exhibit abnormal resonance line flux due to RS effects, even if we observe the integral spectrum from the whole remnant.

    For a better understanding of the possible impact that RS may make on the observed line flux, we performed an MC simulation on the RS processes of O VII resonance line emission in SN 1987A. The fundamental physics of RS and the MC algorithm adopted in our simulation follow \citet{2018ApJ...861..138C} and \citet{2024ApJ...967...99L}. We assumed a ring-like structure for SN 1987A, picturing the dense ER. As demonstrated in Figure \ref{fig:RS_MC}, the geometry of the ring is characterized by an inner radius $R_{\rm i}$, an outer radius $R_{\rm s}$, and a height $h$. The inclination angle $\theta$ is defined as the angle between the ER plane and the line of sight. Given the average radius of the ER as $\sim0''.8$ and the average half-width as $\sim0''.22$ \citep[e.g.,][]{2009ApJ...703.1752R}, we adopted $R_{\rm i}\sim0''.58$ and $R_{\rm s}\sim1''.02$, and assumed the ring height as equal to its width, thereby $h\sim0''.44$. Other parameters (including hydrogen density, O abundance, turbulence velocity, temperature, and ionization parameter) were set as the same as above.

    The MC simulation shows that the maximum optical depth of O VII resonance line along the line of sight, as a function of the inclination angle, can reach $\sim1.7$ when $\theta=0^\circ$ (edge-on) and $\sim0.4$ when $\theta=90^\circ$ (face-on). That means the photon initially {going} along the equatorial plane will have a greater chance to be scattered. Thereby, the line flux will be suppressed along the equatorial plane, while it will be enhanced in polar directions. As a result, the observed G-ratio can be changed due to RS, depending on the inclination angle --- be enhanced by a factor $\sim1.13$ at $\theta=0^\circ$ while being suppressed by a factor $\sim0.95$ at $\theta=90^\circ$ (Figure \ref{fig:RS_MC}). Given the inclination angle of the ER as $\theta\sim47^\circ$ \citep{2005ApJS..159...60S}, the simulation results indicate that the O VII G-ratio will be marginally suppressed by a factor of $\sim0.96$. Therefore, the observed enhancement of G-ratio is unlikely caused by RS effects.
    
    {Similar inferences also work for O VIII Ly$\alpha$ line flux and the resulting Ly$\beta$/Ly$\alpha$ ratio. Despite absolute values being different, the O VIII Ly$\alpha$ line flux will also be suppressed along the equatorial plane while being enhanced in polar directions under the assumption of a ring-like geometry of SN 1987A, just like that for O VII resonance line. At the inclination angle of SN 1987A, we will expect a marginally enhanced O VIII Ly$\alpha$ flux, which will result in a marginally suppressed, rather than enhanced O VIII Ly$\beta$/Ly$\alpha$ ratio (RS makes little effect on O VIII Ly$\beta$ due to its small oscillator strength, as mentioned above). Therefore the observed enhancement of the O VIII Ly$\beta$/Ly$\alpha$ ratio is unlikely due to RS effects, as well.}

    \begin{figure}[ht]
        \centering
        \includegraphics[width=\linewidth]{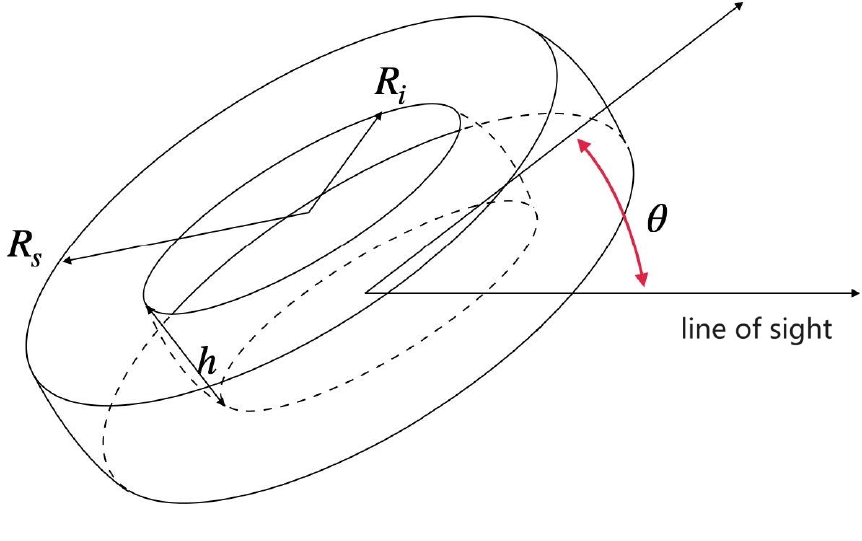}
        \includegraphics[width=\linewidth]{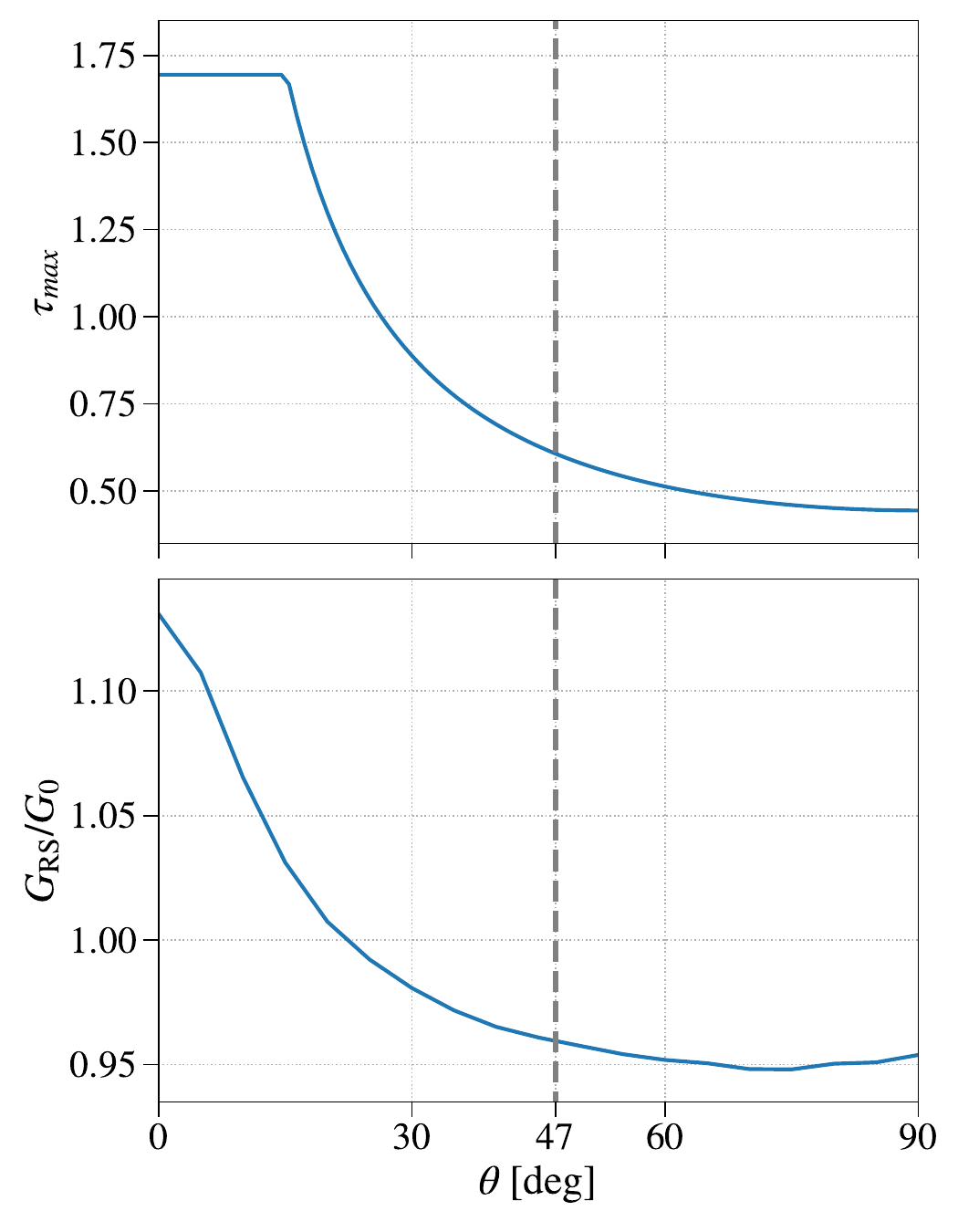}
        \caption{Top: geometry setup adopted for the MC simulation of RS effect in SN 1987A. Middle: maximum RS optical depth as a function of the inclination angle. Bottom: RS-modified G-ratio, relative to the value without RS, as a function of inclination angle. The gray dashed line denotes the inclination angle for the ER of SN 1987A ($\sim47^\circ$).}
        \label{fig:RS_MC}
    \end{figure}
    
    \subsection{Absorption of foreground hot gas}

    \begin{deluxetable*}{ccccccccc}
        \tablecaption{Gaussion absorption fitting result\label{tab:abs}}
        \tablenum{2}
        \tablehead{
        Obs. Date & Age (days) & $\tau_{\rm OVII}$ & $\tau_{\rm OVIII}$ & {$\Delta C$}  & {$\Delta{\rm AIC}$} & {F-test p-value} & {Significance}\tablenotemark{\tiny a} 
        }
        \startdata
        2007 Jan 17 & 7267 & $0.76^{+0.75}_{-0.49}$ & $0.18\pm0.03$ & {$-37.8$} & {$-29.8$} & {$2.5\times10^{-6}$} & {$\gtrsim4.5\sigma$}  \\
        2008 Jan 11 & 7627 & $0.48^{+2.27}_{-0.30}$ & $0.21\pm0.02$ & {$-49.4$} & {$-41.4$} & {$7.9\times10^{-9}$} & {$>5\sigma$}  \\
        2009 Jan 30 & 8012 & $0.53\pm0.16$ & $0.18^{+0.17}_{-0.10}$ & {$-23.3$} & {$-15.3$} & {$6.5\times10^{-4}$} & {$\gtrsim3.5\sigma$} \\
        2009 Dec 11 & 8327 & $0.47\pm0.14$ & $0.20\pm0.02$ & {$-72.4$} & {$-64.4$} & {$1.1\times10^{-11}$} & {$>5\sigma$} \\
        2010 Dec 12 & 8693 & $0.43\pm0.22$ & $0.20\pm0.03$ & {$-48.3$} & {$-40.3$} & {$2.0\times10^{-8}$} & {$>5\sigma$}  \\
        2011 Dec 2 & 9048 & $0.71\pm0.29$ & $0.08\pm0.03$ & {$-17.84$} & {$-9.8$} & {$7.6\times10^{-3}$} & {$\gtrsim2.5\sigma$}  \\
        2012 Dec 11 & 9423 & $0.76\pm0.19$ & $0~(<0.11)$ & {$-10.8$} & {$-6.8$} & {$1.1\times10^{-2}$} & {$\gtrsim2.5\sigma$}  \\
        2014 Nov 29 & 10141 & $0.32~(<0.75)$ & $0.18^{+0.18}_{-0.10}$ & {$-4.7$} & {$+3.3$} & {$0.45$} & {---}  \\
        2015 Nov 15 & 10492 & $0.74\pm0.53$ & $0~(<0.04)$ & {$-11.9$} & {$-5.9$} & {$2.6\times10^{-2}$} & {$\gtrsim2\sigma$}  \\
        2016 Nov 2 & 10845 & $0.49\pm0.22$ & $0~(<0.13)$ & {$-0.3$} & {$+3.7$} & {$0.91$} & {---}  \\
        2017 Oct 15 & 11192 & $0.76~(<1.47)$ & $0.39^{+0.27}_{-0.23}$ & {$-13.1$} & {$-5.1$} & {$3.7\times10^{-2}$} & {$\gtrsim2\sigma$}  \\
        2019 Nov 27 & 11964 & $0.73~(<1.22)$ & $0.05~(<0.28)$ & {$-1.2$} & {$+2.8$} & {$0.60$} & {---}  \\
        2020 Nov 24 & 12328 & $0~(<0.83)$ & $0.36^{+0.19}_{-0.14}$ & {$-8.9$} & {$-0.9$} & {$0.11$} & {$\gtrsim1.5\sigma$}  \\
        2021 Dec 28 & 12727 & $0~(<1.16)$ & $0.11~(<0.18)$ & {$-12.1$} & {$-6.1$} & {$2.5\times10^{-2}$} & {$\gtrsim2\sigma$}  \\
        \enddata
        \tablenotetext{a}{{Approximated from the F-test p-value, only for those with $\Delta{\rm AIC}<0$.}}
    \end{deluxetable*}

    {The X-ray emission of SN 1987A may subject to absorption from the hot, ionized foreground gas, which has not been considered by the {\tt tbvarabs} model adopted in the DEM analysis in \citet{2025inprep}.} 
    {{We note that the underlying physical mechanism here is actually the same as that in RS. As described in Section \ref{sec:rs}, during the RS process, a photon is initially absorbed by an appropriate ion and then promptly re-emitted in a random direction with roughly the same energy. However, when RS occurs along the line of sight but outside the emitting source, the effect of scattering is approximately equivalent to absorption. To differentiate from the RS process occurring within the source as discussed in Section \ref{sec:rs}, we simply refer to it as ``absorption'' in this paper.}}
    
    The resonant absorption of the O VII and O VIII resonance lines may lead to high G ratios and high Ly$\beta$/Ly$\alpha$ ratios. Such hot gas has been observed and constrained in the Galactic halo from both the emission \citep[e.g.,][]{2013ApJ...773...92H,2022PASJ...74.1396U,2023A&A...674A.195P} and the absorption \citep[e.g.,][]{2005ApJ...635..386W,2008ApJ...672L..21Y,2018ApJS..235...28L} perspectives. Additionally, the hot phase ISM in LMC \citep[e.g.,][]{1994ApJ...436L.123S,2002A&A...392..103S} may also contribute to the absorption column density.

    {We examined the absorption scenario by adding two Gaussian absorption components ({\tt gabs}) to the DEM model adopted in \citet{2025inprep}.} We fixed the absorption line centroids (0.5739\,keV for O VII He$\alpha$ resonance line and 0.6535\,keV for O VIII Ly$\alpha$ line) while thawed the line widths $\sigma$ and the absorption strengths $\omega$ {(then the optical depth at the line center is given by $\tau=\frac{\omega}{2\pi\sigma}$)}\footnote{{The optical depth of the foreground hot gas is not expected to be different between different observations. However, since the X-ray flux of SN 1987A keeps changing, it is unable to fix the absorption strength at a same value for all the observations}}. %%\lsun{[response to MM ``{\it - should we consider the Doppler shift associated with the systematic velocity of 87A?}'': The line centroids here are for the foreground absorbing hot gas, therefore have nothing to do with the systematic velocity of SN 1987A.]} 
    We found considerable improvement in C statistics ($\Delta C<-10$) for 10 observations out of 14 {after introducing the Gaussian absorption components}. Particularly, for the first 5 observations (taken from 2007 to 2011, when the O lines were around their maximum fluxes), we got $\Delta C<-20$. {The obtained optical depths and the $\Delta C$ are listed in Table \ref{tab:abs}.}
    
    {In order to better evaluate and demonstrate the significance of the added Gaussian absorption components, we further looked into the Akaike information criterion \citep[AIC,][]{1974ITAC...19..716A} values and performed F-test for each observation. AIC is commonly used as a robust criterion to select the best model that fit to data, where the model with a lower AIC value is preferred. The AIC value can be calculated as:
    \begin{equation}
		{\rm AIC}=2k-2\ln{L}=2k+C,
	\end{equation}
	where $k$ is the number of free parameters and $L$ the likelihood of the model. In our case, a negative $\Delta{\rm AIC}$ suggests that the model with absorption lines offers a better fit to data. As listed in Table \ref{tab:abs}, we obtained $\Delta{\rm AIC}<0$ for 11 out of 14 observations. Especially, all the 7 observations taken in 2007-2012 (when O lines were around their maximum luminosities) result in $\Delta{\rm AIC}<0$, which prefers the hot gas absorption scenario. In addition to AIC, we performed F-test for each observation, and derived the p-value. We also approximated the corresponding sigma levels based on p-values for reference. The F-test p-values, together with the approximated sigma levels, are also listed in Table \ref{tab:abs}. We obtain F-test p-value $<0.05$ for 10 of 14 observations, which may approximately correspond to $\gtrsim2$-sigma significance. For all the 7 observations taken in 2007-2012, we obtained p-value $\lesssim0.01$. The F-test further strengthens the preference of the hot gas absorption scenario. Furthermore, since the spectral fitting and the uncertainty estimation were performed in a Markov Chain Monte Carlo (MCMC) approach \citep[running by the Xspec {\tt chain} command, containing 100,000 effective steps sampling with 40 walkers, for more information of the MCMC simulation, please refer to][]{2025inprep}, we produced and examined the MCMC corner plots of each fit. Figure \ref{fig:opt_dep_corner} provides an example of the corner plots, adopting from the fit to 2009 Jan observation. As seen from the corner plot (e.g., the one for O VII resonance line), despite the large scatterings of the probability distribution functions (PDF) of the absorption strength and the line width as well as the potential degeneracy between them, the resulting PDF of optical depth has a relatively small scattering, indicating a best-fit value significantly larger than zero.}

    \begin{figure*}[ht]
        \centering
        \includegraphics[width=1\linewidth]{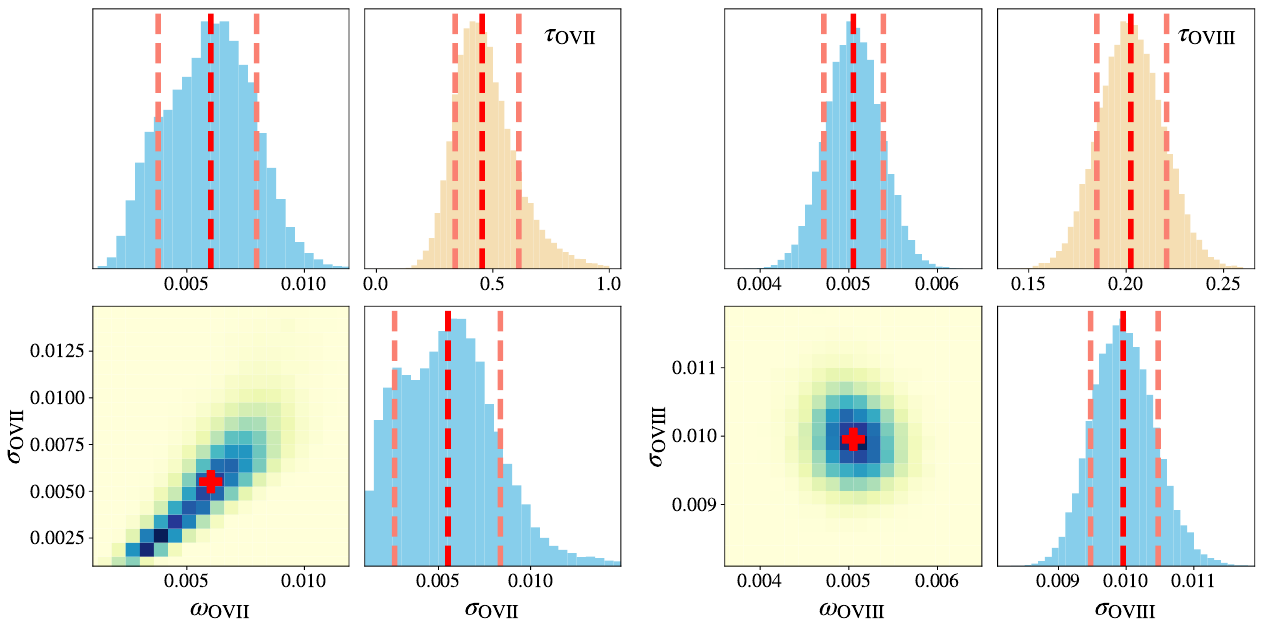}
        \caption{{An example of the MCMC corner plots for the two Gaussian absorption components (O VII resonance line on the left and O VIII Ly$\alpha$ on the right, adopting from the fit to 2009 Jan observation). The top-left and bottom-right panels show the probability distribution functions (PDFs) of the absorption strength and the line width, respectively. The bottom-left panel shows their two-dimensional PDF. The top-right panel shows the PDF of the derived optical depth. The red dashed lines and red crosses denote the best-fit values, while the pink dashed lines denoting the 1-sigma errors.}}
        \label{fig:opt_dep_corner}
    \end{figure*}
    
    Based on the fitting results of these observations, we estimated the average optical depths as $\tau_{\rm OVII}\sim 0.6$ for O VII He$\alpha$ resonance line and $\tau_{\rm OVIII}\sim 0.2$ for O VIII Ly$\alpha$ line.
    The resonant absorption optical depth is formulated as in Eq.\,\ref{eq:1}. Assuming the H-like and He-like O ions share the same ion temperature and turbulence velocity, the ion fraction ratio between H-like and He-like O, $f_{\rm i,OVIII}/f_{\rm i,OVII}$ ($f_{\rm i}\equiv n_{\rm i}/n_{\rm Z}$) can be evaluated as:
    \begin{equation}\label{eq:2}
        \frac{f_{\rm i,OVIII}}{f_{\rm i,OVII}}=\left(\frac{\tau_{\rm OVIII}}{\tau_{\rm OVII}}\right)\left(\frac{E_{\rm OVIII}}{E_{\rm OVII}}\right)\left(\frac{f_{\rm OVIII}}{f_{\rm OVII}}\right)^{-1}
    \end{equation}
    Taking the oscillator strengths of O He$\alpha$ resonance line and Ly$\alpha$ line as $f_{\rm OVII}=0.72$ and $f_{\rm OVIII}={0.42}$\footnote{The oscillator strengths of the two O Ly$\alpha$ lines are 0.28 for Ly$\alpha_1$ (2p$^2$P$_{3/2}\rightarrow$1s$^2$S$_{1/2}$) and 0.14 for Ly$\alpha_2$ (2p$^2$P$_{1/2}\rightarrow$1s$^2$S$_{1/2}$). { Given the Ly$\alpha$1 and Ly$\alpha$2 lines are rather close in energy, they are not absorbed separately --- photons from Ly$\alpha$1 or Ly$\alpha$2 can be absorbed by either line}. Therefore, we take the {sum of their oscillator strengths}.}, we got $f_{\rm i,OVIII}/f_{\rm i,OVII}\sim{0.65}$. Assuming the hot gas is under the CIE state, the H-like/He-like O ion fraction ratio gives {an} electron temperature $kT_{\rm e}\sim{0.15}$\,keV. This temperature is consistent with the hot gas in the Galactic halo, for which {other authors} found a temperature $\sim0.15$--$0.22$\,keV \citep[e.g.,][]{2013ApJ...773...92H,2022PASJ...74.1396U,2023A&A...674A.195P}. Assuming the ion temperature is equal to the electron temperature and a turbulence velocity of 100\,km\,s$^{-1}$, we further estimated the column density of the oxygen based on Eq.1 as $N_{\rm O}\sim{0.5}\times10^{16}$\,cm$^{-2}$. The oxygen column density we found is comparable with those suggested by previous studies on Galactic halo X-ray absorption lines \citep[$\sim10^{15}$--$10^{16}$\,cm$^{-2}$, e.g.,][]{2005ApJ...635..386W,2008ApJ...672L..21Y,2018ApJS..235...28L}.

    \begin{deluxetable*}{cccccc}
        \tablecaption{{Hot gas absorption corrected line ratios and O abundances}\label{tab:abs_corr}}
        \tablenum{3}
        \tablehead{
        {Obs. Date} & {Age (days)} & {G-ratio}\tablenotemark{\tiny a} & {Ly$\beta$/Ly$\alpha$}\tablenotemark{\tiny a} & {O}\tablenotemark{\tiny b} ({this work}) & {O}\tablenotemark{\tiny b} ({Sun+24})
        }
        \startdata
        {2007 Jan 17} & {7267} & {$0.53^{+0.13}_{-0.12}$} & {$0.16\pm0.02$} & {$0.219\pm0.010$} & {$0.177^{+0.016}_{-0.010}$} \\
        {2008 Jan 11} & {7627} & {$0.67\pm0.15$} & {$0.15\pm0.01$} & {$0.328\pm0.014$} & {$0.249\pm0.017$} \\
        {2009 Jan 30} & {8012} & {$0.77\pm0.15$} & {$0.18\pm0.01$} & {$0.279^{+0.011}_{-0.012}$} & {$0.238^{+0.022}_{-0.010}$} \\
        {2009 Dec 11} & {8327} & {$0.61\pm0.14$} & {$0.18\pm0.02$} & {$0.343\pm0.012$} & {$0.324^{+0.048}_{-0.011}$} \\
        {2010 Dec 12} & {8693} & {$0.63^{+0.17}_{-0.14}$} & {$0.18\pm0.02$} & {$0.538\pm0.022$} & {$0.450^{+0.087}_{-0.012}$} \\
        {2011 Dec 2} & {9048} & {$0.71\pm0.17$} & {$0.21\pm0.02$} & {$0.343\pm0.015$} & {$0.283^{+0.024}_{-0.023}$} \\
        {2012 Dec 11} & {9423} & {$0.64^{+0.18}_{-0.17}$} & {$0.23\pm0.02$} & {$0.383\pm0.010$} & {$0.368^{+0.079}_{-0.025}$} \\
        {2014 Nov 29} & {10141} & {$0.79\pm0.22$} & {$0.19\pm0.02$} & {$0.440\pm0.022$} & {$0.373^{+0.041}_{-0.033}$} \\
        {2015 Nov 15} & {10492} & {$0.57\pm0.25$} & {$0.20\pm0.03$} & {$0.269\pm0.007$} & {$0.336^{+0.071}_{-0.024}$} \\
        {2016 Nov 2} & {10845} & {$0.47^{+0.22}_{-0.17}$} & {$0.21\pm0.03$} & {$0.253\pm0.010$} & {$0.239^{+0.20}_{-0.29}$} \\
        {2017 Oct 15} & {11192} & {$0.69\pm0.26$} & {$0.20\pm0.02$} & {$0.255^{+0.18}_{-0.17}$} & {$0.190^{+0.029}_{-0.013}$} \\
        {2019 Nov 27} & {11964} & {$0.76~(<1.55)$} & {$0.26\pm0.06$} & {$0.406^{+0.041}_{-0.025}$} & {$0.396^{+0.138}_{-0.023}$} \\
        {2020 Nov 24} & {12328} & {$0.29^{+0.27}_{-0.22}$} & {$0.15\pm0.03$} & {$0.350\pm0.020$} & {$0.268^{+0.041}_{-0.014}$} \\
        {2021 Dec 28} & {12727} & {$0.76~(<1.64)$} & {$0.18\pm0.04$} & {$0.189^{+0.013}_{-0.008}$} & {$0.182^{+0.033}_{-0.012}$} \\
        \enddata
        \tablenotetext{a}{{Hot gas absorption corrected (i.e., intrinsic) line ratios.}}
        \tablenotetext{b}{{O abundance in units of solar abundance \citep[][]{2000ApJ...542..914W}.}}
    \end{deluxetable*}

    {Neglecting foreground hot gas absorption in previous studies led to an underestimation of the O VII resonance line and O VIII Ly$\alpha$ line fluxes, and thus an overestimation of the intrinsic O VII G ratios and O VIII Ly$\beta$/Ly$\alpha$ ratios. By incorporating the best-fit Gaussian absorption line components into the ``{\tt nlapec} + {\tt gauss}'' model as described in Section 3, we measured the hot-gas-absorption-corrected O line fluxes. The intrinsic G-ratios and Lyb/Lya ratios were then calculated and listed in Table \ref{tab:abs_corr}. For most of the observations (especially those with significant absorption line detections), we obtained corrected O VII G-ratios in the range of 0.5--0.8 and corrected O VIII Ly$\beta$/Ly$\alpha$ ratios in the range of 0.15--0.20. An intrinsic O VII G-ratio $\sim0.5$--$0.8$ is consistent with the expectation of NEI plasma with a temperature of $\sim0.3$--$1$\,keV and an ionization parameter of $10^{11}$--$10^{12}$\,cm$^{-3}$\,s, as seen in Section \ref{sec:nei} and Figure \ref{fig:line_ratio_compared}. On the other hand, an intrinsic O VIII Ly$\beta$/Ly$\alpha$ ratio of $\sim0.15$--$0.20$ seems to be still higher than what we expect for NEI/CIE plasma. However, considering a potential contribution of Fe XVIII F6 as (on average) $\sim25\%$ to the measured Ly$\beta$ flux (see the discussions in Section \ref{sec:result}), the real Ly$\beta$/Ly$\alpha$ ratio could also be consistent with the NEI/CIE plasma interpretation. Given the fact that the physical parameters (e.g., temperature and ionization parameter) of the hot plasma in SN 1987A keep varying, the intrinsic O line fluxes and ratios are expected to be varying as well. Specifically, as shown in Figure \ref{fig:line_ratio_compared}, the intrinsic O VII G ratio is expected to be decreasing, while O VIII Ly$\beta$/Ly$\alpha$ to be increasing with an increasing average temperature. Also, we note that the potential contribution of Fe XVIII F6 emission to the measured O VIII Ly$\beta$ flux, and thus the measured Ly$\beta$/Ly$\alpha$ ratio, are expected to be increasing with increasing temperature. However, due to the large uncertainties, we could not address a significant time evolution trend of the measured intrinsic line ratios.}
    
    {Neglecting foreground hot gas absorption also led to an underestimation of the O abundance {in the previous studies}. For the observations {taken in 2007--2010 (with $\gtrsim3\sigma$ detection of the absorption components}, the {original} DEM model gives O abundances ranging in $\sim0.17$--$0.45$, with an average value $\sim0.29$ \citep{2025inprep}. After {incorporating} the absorption components, the obtained best-fit O abundances show an average increase of $\sim20\%$, ranging in $\sim0.22$--$0.53$ with an average value $\sim0.34$. On the other hand, the N/O ratio (by number of atoms) in this case is $\sim1.2$, which is consistent with the values obtained from optical studies \citep[e.g., $\sim1.1$--$1.5$,][]{1996ApJ...464..924L,2010ApJ...717.1140M}.

\section{Conclusion}\label{sec:conclusion}

{In this work, we took a further look into the residuals left in the DEM modeling of SN 1987A by \citet{2025inprep}, focusing on the O lines. We revised and updated the O line fluxes on the basis of \citet{2021ApJ...916...41S}. We found a high O VII G-ratio $\gtrsim1$ and a high O VIII Ly$\beta$/Ly$\alpha$ ratio $\gtrsim0.2$ in SN 1987A. In order to explore {their} physical origin, we performed a detailed investigation into four possible scenarios, i.e., the NEI effects, {CX}, {RS}, and the hot gas absorption. None of the NEI, CX, or RS can fully explain the observed O line ratios, while the spectral fitting can be considerably improved by adding two Gaussian absorption components at O VII resonance line and O VIII Ly$\alpha$ line. 
%As a conclusion, we suggest} the high G-ratio and Ly$\beta$/Ly$\alpha$ ratio in SN 1987A are most likely to be caused by the absorption of the foreground hot gas. 
{We obtained the optical depths as $\tau_{\rm OVII}\sim0.6$ for O VII resonance line and $\tau_{\rm OVIII}\sim0.2$ for O VIII Ly$\alpha$ line.} The estimated temperature ($kT_{\rm e}\sim{0.15}$\,keV) and column density ($N_{\rm O}\sim{0.5}\times10^{16}$\,cm$^{-2}$) of the absorbing gas is consistent with the hot Galactic X-ray halo. {We suggest, therefore, that the high G-ratio and Ly$\beta$/Ly$\alpha$ ratio in SN 1987A are most likely caused by the absorption of the foreground hot gas.}
{The O abundance of SN 1987A might be underestimated in previous studies due to the neglecting of the foreground hot gas absorption. We found an average increase of $\sim20\%$ in the best-fit O abundances after adding the absorption components, which leads to an N/O ratio $\sim1.2$ (by number of atoms).}

However, we note that while the hot gas absorption is likely playing the major role, all the possible mechanisms mentioned above may simultaneously contribute to the modification of the line ratios. 
We also note that already a few LMC SNRs have been found to exhibit unusual O line ratios, such {as} DEM L71 \citep{2003A&A...406..141V}, N49 \citep{2020ApJ...897...12A}, N132D \citep{2020ApJ...900...39S}, and J0453.6$-$6829 \citep{2022PASJ...74..757K}. Even though for most of them, the abnormal line ratios have been attribute to CX or RS effects, the absorption of the hot Galactic halo gas may contribute as well. {In the near future, spatially-resolved high-resolution X-ray spectroscopic studies with, e.g., XRISM, HUBS, and LEM, may help to better constrain the contributions of CX, RS, and hot gas absorption in these remnants.}

\begin{acknowledgments}

	L.S.\ and Y.C.\ acknowledge the NSFC fundings under grants 12173018, 12121003, and 12393852. L.S.\ acknowledges the support from Jiangsu Funding Program for Excellent Postdoctoral Talent (2023ZB252). P.Z.\ thanks the support from NSFC grant No.\ 12273010.
    S.O., E.G., and M.M. acknowledge financial contribution from the PRIN 2022 (20224MNC5A) - ``Life, death and after-death of massive stars'' funded by European Union – Next Generation EU, and the INAF Theory Grant ``Supernova remnants as probes for the structure and mass-loss history of the progenitor systems''.
    {We are grateful for valuable suggestions from John Raymond.}

\end{acknowledgments}

\software
{XSPEC} \citep{1996ASPC..101...17A}, SPEX \citep{1996uxsa.conf..411K}, SAS \citep{2004ASPC..314..759G}


\bibliography{LS_ref}{}
\bibliographystyle{aasjournal}



\end{document}